\documentclass[sigconf]{acmart}

\usepackage{booktabs} 

\usepackage[linesnumbered,ruled]{algorithm2e}
\usepackage{soul}
\usepackage{comment}

\copyrightyear{2019} 
\acmYear{2019} 
\setcopyright{acmlicensed}
\acmConference[WSDM Cup workshop '19]{12th ACM International Conference on Web Search and Data Mining}{February 15, 2019}{Melbourne, Australia}
\acmBooktitle{12th ACM International Conference on Web Search and Data Mining, February 15, 2019, Melbourne, Australia}
\acmPrice{15.00}
\acmDOI{}
\acmISBN{}

\begin{document}
\title[Skip prediction using boosting trees]{Skip prediction using boosting trees based on acoustic features of tracks in sessions}


\author{Andres Ferraro}
\affiliation{%
  \institution{Music Technology Group - Universitat Pompeu Fabra}
  \streetaddress{Roc Boronat 138}
  \city{Barcelona}
  \state{Spain}
  \postcode{08018}
}
\email{andres.ferraro@upf.edu}

\author{Dmitry Bogdanov}
\affiliation{%
  \institution{Music Technology Group - Universitat Pompeu Fabra}
  \streetaddress{Roc Boronat 138}
  \city{Barcelona}
  \state{Spain}
  \postcode{08018}
}
\email{dmitry.bogdanov@upf.edu}

\author{Xavier Serra}
\affiliation{%
  \institution{Music Technology Group - Universitat Pompeu Fabra}
  \streetaddress{Roc Boronat 138}
  \city{Barcelona}
  \state{Spain}
  \postcode{08018}
}
\email{xavier.serra@upf.edu}

\renewcommand{\shortauthors}{A. Ferraro et al.}

\begin{abstract}
The Spotify Sequential Skip Prediction Challenge focuses on predicting if a track in a session will be skipped by the user or not. In this paper, we describe our approach to this problem and the final system that was submitted to the challenge by our team from the Music Technology Group (MTG) under the name ``aferraro''. This system consists in combining the predictions of multiple boosting trees models trained with features extracted from the sessions and the tracks. The proposed approach achieves good overall performance (MAA of 0.554), with our model ranked 14th out of more than 600 submissions in the final leaderboard.
\end{abstract}

%
%



\maketitle
\section{Introduction}

The Spotify Sequential Skip Prediction Challenge \cite{brost2019music} consists in building a system for predicting the tracks that will be skipped in a test set of user listening sessions. The evaluation is done taking into account the position of the tracks in the session. For each session in the test set the first half of the tracks includes the information whether those tracks were skipped or not. Participants must predict skipped tracks for the other half.

A track is considered skipped when a user did not listen to the entire track, but the dataset contains more refined skip annotations for each track:
\begin{itemize}
\item skip\_1: if the track was played very briefly 
\item skip\_2: if the track was played only briefly 
\item skip\_3: if most of the track was played
\end{itemize}
The ground-truth for evaluation is based only on the information of skip\_2.

The challenge is organized in collaboration with the Spotify music streaming service, who provided the dataset for the challenge. The training set contains nearly 130 million listening sessions with some extra information and the test set contains 30 million sessions. The maximum session length is 20 tracks. Overall, these sessions cover nearly 4 million different music tracks and the organizers provided acoustic features and metadata for these tracks as an additional data to be exploited in the challenge. It was also allowed to use other sources of information. 

With the transformation of the digital music industry, music streaming services are increasingly prevailing 
and it is important to understand music consumption on such platforms. According to reports by Spotify~\cite{lamere}, a track has a 24\% 
likelihood of being skipped in the first 5 seconds and 35\% 
of being skipped before 30 seconds. Reducing such a skipping behaviour of users may be a valuable strategy 
in order to improve engagement with the platform and attract new users.

Automatic recommendations play an important role in music consumption on streaming services and this topic is widely studied in the literature \cite{schedl2015music, schedl2018current}. One common approach is to use the implicit feedback information, such as the interactions between the users and the tracks (e.g., the number of times a user plays a track), to generate recommendations. Some works also use the information about skipping behaviour to improve their systems and either train the systems or measure their performance using such data~\cite{pampalk2005dynamic, bonnin2015automated, moling2012optimal, hu2011nextone}.
Existing research on music recommender systems has considered a number of related tasks, including Automatic Playlist Generation and Automatic Playlist Continuation. The former consists in automatic creation of a sequence of tracks with some common characteristic or intention, while the latter considers inference of those properties from the existing playlists for their automatic continuation. 

Recently Spotify has also organized the RecSys Challenge focused on Automatic Playlist Continuation~\cite{RecSysChallenge2018}. Some of the best performing solutions incorporated track and playlist features in addition to track-playlist relations into their models~\cite{zamani2018analysis}. 
Our team (``cocoplaya'') proposed a hybrid system that combines acoustic features from the tracks extracted with Essentia,\footnote{\url{http://essentia.upf.edu}} an open-source library for audio analysis for music information retrieval applications~\cite{bogdanov2013essentia}. We combined the acoustic features together with other features extracted from the playlists and the interactions between tracks and playlists. The proposed solution 
gained the 4th position in the creative track of the challenge~\cite{ferraro2018automatic}. The solution of the winner team \cite{volkovs2018two} uses a model based on boosting trees to improve the recommendation based on some acoustic features and other features extracted from the playlist and the tracks. 

Since the two problems are similar, for the Sequential Skip Prediction Challenge we proposed a similar solution. We extract features from listening sessions and tracks and train boosting trees using these features to identify which tracks are skipped by the users. Remarkably, the solutions for this challenge are computationally demanding and must be memory efficient due to the size of the dataset. This has limited us in the considered solutions.

\section{Features extracted from tracks and sessions}
In this section we describe the features used by our system. 
As it has been shown that it is possible to reduce the number of skips using audio similarity~\cite{pampalk2005dynamic}, most of the features used in our solution are acoustic features. We also use features that describe the time at which a particular session happened because it has been noted that there is a significant difference in the number of skips depending on this information~\cite{lamere}.

Our goal is to design a system that predict skips for a track given previous history of the session. To this end, our system uses the acoustic features of the track together with the features characterizing the session as an input. 

In the challenge, 
the information about skips is provided only for the first half of each session in the test dataset and the goal is to predict the skips in the second half.
We perform a similar split for the training set and use the tracks in its second half as training examples annotated by skipping behavior (the \textit{skipping behaviour set}).
For each example, we compute the acoustic features of the corresponding track together with the features characterizing the previous history of the session it belongs. For the latter we use the first half of the session split (the \textit{session history}). A total of 63 features described below were computed for each training example.


%



\subsection{Track features}
Each track was characterized by the following 16 features available via the Spotify API:\footnote{\url{https://developer.spotify.com/documentation/web-api/reference/tracks/get-audio-features/}}
popularity, acousticness, beat strength, bounciness, danceability, mean dynamic range, energy, flatness, instrumentalness, liveness, loudness, mechanism, tempo, organism, speechiness and valence.

\subsection{Session features}
Here, we are interested to characterize a typical track that is skipped or listened entirely in a session. To this end, we compute the average of the track features in the session history for both skipped and listened tracks resulting in the additional 32 features. 
In addition, the following session features are computed based on the last track in the first half of the session: premium, shuffle, hour, day and month.

The binary feature ``premium'' is true when the user associated with the session has a premium account. This feature is important since non-premium users have a restricted number of skips. The feature ``shuffle'' indicates if the user was listening to tracks in a shuffle mode. The other three features represent the time when the user listened to the songs. This is also important since the behaviour of the users is different depending the the moment of the day.
Finally, we also computed the ratio of the skipped tracks among all tracks in the session history with respect to the two variants of skips (skip\_1 and skip\_2).

\subsection{Session-Track features}

Following the leading approach in the RecSys challenge~\cite{volkovs2018two} we also computed some features that combine the information from the track and the session history at the same time. 
Using those features we want to measure if the track is more similar to the tracks that were previously skipped by the user or to the ones that were listened.
To this end, we compute the average difference between the track and the skipped/listened tracks in the session history in terms of 
duration, year and popularity as well as the average value of the dot product between the acoustic vectors of the tracks.


In addition to the aforementioned acoustic track features, Spotify provides a 7-dimensional acoustic vector \cite{NIPS2013_5004} representing each track. For each training example we compute an average dot product between the acoustic vector of the corresponding track and each of the tracks in the session history.

\section{Model}

Once all the features were computed for training examples  we used them to train the model. Following~\cite{volkovs2018two} we built a model based on boosting trees to predict the skips, using the same library (xgboost\footnote{\url{https://xgboost.readthedocs.io}}). It was not possible to train the model loading the training set in batches. The xgboost library recommends to use Apache Spark when the training set is too large, but we did not have the required infrastructure. We therefore decided to divide the data 
and train multiple position-dependent models that predict a skip at a particular position in the playlist. We split the training examples into 10 subsets according by their position index in the skipping behaviour set (from 1 to 10). In total we trained 10 models.

To decide the best model parameters to use we selected a random sample (5\%) of the data. Combinations of the following parameter values were tested:
\begin{itemize}
\item eta: $0.1$, $0.2$ ,$0.3$
\item max\_depth: $6$, $10$, $15$
\item colsample\_bytree: $0.8$, $1.0$
\item subsample: $0.8$, $0.9$ ,$1.0$
\end{itemize}
The value of num\_boost\_round used was set to 200 and the best parameters found were  eta=$0.3$, max\_depth=$15$, subsample=$1.0$ and colsample\_bytree=$1.0$. This parametrization resulted in a model with the AUC of 0.732.

\section{Results}
The test set contains user sessions in each of which only the first half of the tracks is annotated by user actions (skips). The task is to predict the actions for the rest of the tracks, for which the organizers have the ground truth. We extracted the features for all these tracks and sessions in the same way as when training our models. 

We combined the trained models in different ways to produce the best results. In this section we explain our submitted solutions and report the evaluation results.

\subsection{Metrics and evaluation}

For the evaluation the organizers used the average accuracy to evaluate every session:
\begin{displaymath}
  AA = \frac{\sum_{i=1}^{T} A(i)L(i)}{n}
\end{displaymath}
where 
$n$ is the number of tracks in the sequence for which the skips should be predicted for a given session,
$A(i)$ is the accuracy at position $i$ of the track sequence, and
$L(i)$ is a Boolean indicator for if the skip prediction at the position $i$  was correct.
Results were averaged across sessions to compute the mean average accuracy (MAA).

In order to evaluate the systems, participants were requested to submit 
predictions of skips (0 or 1) for each session in the test set. Only five submissions could be done by each team per day of the challenge.
During the challenge, the organizers published a leaderboard with the positions of each team with respect to their MAA score. In the case multiple teams head the same MAA value, the accuracy at the first position was used for a tiebreak. The organizers used only 50\% of the challenge test set to calculate the scores during the challenge and used a private set for the final evaluation afterwards.

\subsection{Submitted solutions}
In this section we describe our solutions used to generate the submissions and compare their performance. 

A sequence of test tracks $\{t_1,...,t_n\}, n\leq10$ is given as an input to the system one at a time to predict the skip behavior in a session. 
We have trained 10 position-dependent prediction models $M_j$, one for each possible track position $j \in [1,10]$. 
Given a track $t$ represented by its track and session features, every model $M_j$(t) outputs a value between 0 and 1, where values closer to 1 mean that is more probable to be a skip.
We use the combination of these models to predict the skip action $S(t_i) \in \{0,1\}$ for each track $t_i$ at the position $i$ in a test sequence.


The simplest approach is to predict the user action for a track $t_i$ using only the corresponding model $M_i$ trained for the same position index. This resulted in a performance below the highest baseline.
Such an approach naively consider the outcomes for all tracks in the test set independently instead of considering the session as a sequence of action. It may be important to know user action at the previous track to predict the next action. Therefore, we explored different ways of combining the predictions from position-dependent models $M_i$ in combination with the last known user action in the session history (the \textit{last user action}), denoted as $S(t_0)$. By default we assign $S(t_0)=1$ if the last user action was a skip, $S(t_0)=0$ otherwise.

Table~\ref{tab:submission} summarizes the results for all considered solutions described below:


\begin{itemize}
\item \textbf{Solution 1:} As the simplest solution, we only use the output of the corresponding position-dependent model $M_i(t_i)$ for each track $t_i$ in the session. 

\item \textbf{Solution 2:} We average the last user action $S(t_0)$ with the prediction of Solution~1:
\begin{displaymath}
 S(t_i) = 0.5*M_i(t_i) + 0.5*S(t_0)
\end{displaymath}

\item \textbf{Solution 3:} Same as Solution 2 but using $S(t_0)=0.6$ when the last track was skipped and $S(t_0)=0.4$ otherwise.  With this change we give less weight to the last known action of the user.

\item \textbf{Solution 4:} For a track $t_i$ at the position $i$ in the test sequence, we average model predictions:

 \begin{displaymath}
 Q(t_i) = \frac {1}{i} * \sum_{j=1}^{i} M_j(t_i)
\end{displaymath}

and then compute the final prediction:

\begin{displaymath}
 S(t_i) = 0.5*Q(t_i) + 0.5*S(t_0) 
\end{displaymath}

\item \textbf{Solution 5:} Same as Solution 4 but $S(t_0)=0.6$ if the track was skipped, $S(t_0)=0.4$ otherwise. 

\item \textbf{Solution 6:} For this solution first we average two groups of model predictions Q and W:

 \begin{displaymath}
 Q(t_i) = \frac {1}{5} * \sum_{j=1}^{5} M_j(t_i) , \quad  
 W(t_i) = \frac {1}{5} * \sum_{j=6}^{10} M_j(t_i) 
\end{displaymath}

and then compute the final prediction as:
\begin{displaymath}
S(t_i) =
\begin{cases}
    0.4*S(t_0) + 0.4*Q(t_i) + 0.2*W(t_i),& \text{if } i\leq 5\\
    0.2*S(t_0) + 0.5*Q(t_i) + 0.3*W(t_i),& \text{otherwise}
\end{cases}
\end{displaymath}

\item \textbf{Solution 7:} Same as Solution 6 but using different weights for the case when $i > 5$: 

\begin{displaymath}
 S(t_i) = 0.4*S(t_0) + 0.3*Q(t_i) + 0.3*W(t_i)
\end{displaymath}

\item \textbf{Solution 8:} The predictions are computed as:
\begin{displaymath}
 S(t_i) = 0.5*\frac {1}{n}\sum_{j=1}^{n} M_j(t_i) + 0.5*S(t_{i-1})
\end{displaymath}
Note that in the case where we calculate $S(t_1)$, we use the default values for the last user action $S(t_0)$.


\item \textbf{Solution 9:} Again, we use all 10 position-dependent models $M_j$, but we give more weight to the models that are closer to the position of the track $t_i$ in a linear scale. We also consider the action of the last track in the session history $S(t_0)$ with a weight according to its distance to $t_i$.\footnote{Refer to the source code for more details.}

%

\item \textbf{Solution 10:} This is the same as Solution 9 but in this case the weight of $S(t_0)$ does not depend on the distance to the track $t_i$ but is set to a fixed weight of 20\%.

\item \textbf{Solution 11:} This is the same solution as Solution 8 but we give a weight of 0.4 to $S(t_{i-1})$.

\item \textbf{Solution 12:} This is the same as Solution 9 but in this case we assign an exponential weight to each model according to the distance to $t_i$.
\end{itemize}

\begin{table}
  \caption{Evaluation results for the submitted solutions.}
  \label{tab:submission}
  \begin{tabular}{lcc}
\toprule
Solution&MAA&First Prediction Accuracy\\
\midrule
Solution 1&0.529&0.693\\
Solution 2&0.535&0.739\\
Highest baseline&0.537&0.742\\
Solution 7&0.539&0.743\\
Solution 6&0.542&0.743\\
Solution 5&0.543&0.724\\
Solution 3&0.546&0.724\\
Solution 11&0.549&0.742\\
Solution 12&0.549&0.729\\
Solution 4&0.550&0.742\\
Solution 8&0.550&0.742\\
Solution 10&0.551&0.735\\
Solution 9&\textbf{0.554}&0.735\\
\bottomrule
\end{tabular}
\end{table}

As it can be seen in Table~\ref{tab:submission}, the best performance was achieved by Solution 9 (MAA of 0.554), being close to the performance of Solution 10 (MAA 0.551). All solutions performed better then the best baseline (MAA 0.537), except for Solution 1. This demonstrates the importance of employing information about previous user actions for better predictions.

\section{Conclusions and future work}
In this paper we describe our solutions for the Spotify Sequential Skip Prediction Challenge, 
the best of which achieved the 14th position in the leaderboard. Our approach consists in combining the predictions of multiple models based on boosting trees trained on features characterizing tracks and sessions.

One advantage of our solution is that it is possible to incorporate more features into the model and we expect that including more relevant information will improve the system. Given the time constraints of the challenge, we were not able to evaluate all combinations of acoustic features and other metadata as we planned. We expect that other acoustic features can improve the performance of our solution and this idea will be addressed in the future work. Our current solutions are only limited to the features provided by the Spotify API. We will consider using high-level music features available in Essentia audio analysis library, such as genre and mood annotations. We also expect that extracting artist-related features for tracks and sessions can provide a better result.

Another improvement could be by using a matrix factorization approach to get a representation of the sessions and the tracks based on their interactions~\cite{volkovs2018two, ferraro2018automatic}. 
However, such a solution will be very computationally demanding due to the size of the dataset. 

The code of our submissions is open-source\footnote{\url{https://github.com/andrebola/skip-challenge-wsdm}} and we encourage other researchers to experiment with our system and combine it with other solutions.

\begin{acks}
This research has been supported by Kakao Corp., 
and partially funded by the European Unions Horizon 2020 research and innovation programme under grant agreement No 688382 (AudioCommons) and 
the  Ministry of Economy and Competitiveness of the Spanish Government (Reference: TIN2015-69935-P). 
\end{acks}

\bibliographystyle{ACM-Reference-Format}
\bibliography{sample-sigconf}

\end{document}